\documentclass[12pt]{article}
\usepackage[utf8]{inputenc}
\usepackage[left=1in, right=1in, top=1in]{geometry}
\usepackage{amsmath}
\usepackage{amsfonts}
\usepackage{clipboard}
\usepackage{cite}
\usepackage[colorlinks=true, linkcolor = blue, citecolor = blue, urlcolor = blue,linktocpage=true]{hyperref}
\usepackage{amssymb}
\usepackage{graphicx}
\usepackage{float}
\usepackage{caption}

\title{Accordiohedra as Positive Geometries for Generic Scalar Field Theories}
\author{Mrunmay Jagadale, Nikhil Kalyanapuram, Aneesh P. B.}
\date{April 2019}

\begin{document}


\begin{centering}

{\LARGE Accordiohedra as positive geometries for generic scalar field theories \par}

 \vspace{1cm}
P. B. Aneesh, Mrunmay Jagadale, Nikhil Kalyanapuram
  \vspace{0.8cm}

 \vspace{0.8cm}

\begin{minipage}{.9\textwidth}\small  \begin{center}
Chennai Mathematical Institute, H1 SIPCOT IT Park, \\ Kelambakkam, Tamil Nadu, 603103, India \\
  \vspace{0.8cm}
{\tt   aneeshpb, jmrunmay, nikhilkaly@cmi.ac.in}
\\ $ \, $ \\
\end{center}
\end{minipage}

\end{centering}

%
%
%
%
%

\begin{abstract}
We build upon the prior works of \cite{Arkani-Hamed:2017mur,Banerjee:2018tun,Raman:2019utu} to study tree-level planar amplitudes for  a massless scalar field theory with polynomial interactions. Focusing on a specific example, where the interaction is given by $\lambda_3\phi^{3}\ +\lambda_4 \phi^{4}$, we show that a specific convex realization of a simple polytope known as the accordiohedron in kinematic space is the positive geometry for this theory. As in the previous cases, there is a unique planar scattering form in kinematic space, associated to each positive geometry which yields planar scattering amplitudes. 

\end{abstract}

\newpage
\tableofcontents

\section{Introduction}

The quest for analyzing the structural aspects of the S-matrix of quantum field theories continues to explore new dimensions. One of the beautiful recent developments in this quest is the ``Amplituhedron program", in which the fundamental object of interest is a positive geometry (in fact, a convex polytope) called the Amplituhedron \cite{Arkani-Hamed:2013jha, Arkani-Hamed:2013kca}. In the context of  planar amplitudes in ${\cal N} = 4$ Super Yang-Mills theory, the amplituhedron is embedded in an ambient space such as the momentum twistor space, and a unique form on the ambient space which is determined completely by the amplituhedron, determines the S matrix . This formulation, if successful for a wider class of theories, has far reaching ramifications in understanding scattering amplitudes; for properties like unitarity and locality turn out to be consequences of the topology and geometry of the polytope. For a number of striking developments in the amplituhedron program, we refer the reader to \cite{Arkani-Hamed:2014dca},\cite{Arkani-Hamed:2017vfh},\cite{Arkani-Hamed:2018rsk}.

In \cite{Gao:2017dek,Arkani-Hamed:2017mur} the amplituhedron program was extended from the world of super-symmetric quantum field theories to tree-level scattering amplitudes of bi-adjoint $\phi^{3}$ theory. The amplitude of the theory was understood in terms of a certain canonical form associated to a positive geometry embedded in the kinematic space of mandelstam invariants. This positive geometry turned out to be a specific convex realization of a well known combinatorial object known as the associahedron. This new understanding of the amplitude as a differential form on very special geometries has far reaching ramifications, as it sheds new light on color-kinematics duality, geometric understanding of recursion relations and the Cachazo-He-Yuan (CHY) formula \cite{Cachazo:2013iea}.

In \cite{Banerjee:2018tun,Raman:2019utu} the authors tried to locate the positive geometries associated to tree-level planar amplitudes in massless scalar field theories with $\phi^{p}\ (p\ \geq\ 4)$ interactions. Although there was no single polytope whose associated canonical form determined the amplitude, a set of polytopes encoded the information about scattering amplitudes and hence the role of amplituhedron for these theories was played by the union over this set of polytopes. 

Combining the results of \cite{Arkani-Hamed:2017mur, Banerjee:2018tun, Raman:2019utu}, an interesting picture emerged regarding positive geometries associated to monomial scalar interactions. Namely, the positive geometries associated to such interactions belonged to a family of polytopes collectively called accordiohedra \cite{manneville2019geometric, padrol2019associahedra} . Accordiohedra are a (infinite) family of simple polytopes which contain associahedra, Stokes polytopes \cite{baryshnikov2001stokes,chapoton2015stokes} as well as those polytopes which are combinatorially built out of non-crossing dissections (see section \ref{accordiohedron} for details). 

In this paper we attempt to extend the amplituhedron program further by considering polynomial scalar interactions of the form $\sum_{n=3}^{N}\lambda_{n}\phi^{n}$ . For concreteness, we analyze $\lambda_{3}\phi^{3}\ +\ \lambda_{4}\phi^{4}$ potential and show that the convex realizations of the combinatorial polytopes which belong to the accordiohedron family are indeed the positive geometries for corresponding scattering amplitudes. The paper is organized as follows.

In section \ref{accordiohedron}, we propose a positive geometry of a general planar scalar field theory and obtain a scattering form associated with it. An arbitrary scalar field theory is defined by a tower of interaction vertices, starting with a three-point coupling. We will see that a polytope known as the accordiohedron supplies a suitable candidate for the corresponding positive geometry. In the case of pure cubic or pure quartic interactions, we will see that it reduces to the associahedron and Stokes polytope, respectively. In section \ref{sixpoint}, we will look at an example of accordiohedron in the case of six-point with two cubic vertices and a quartic vertex. 

To get the scattering amplitude we need to embed the accordiohedron in kinematic space (space spanned by Mandelstam variables) and pull-back the scattering form associated with the accordiohedron onto the embedded accordiohedra. In section \ref{embedding} we give an embedding of accordiohedron in the kinematic space which, unlike the embedding of Stokes polytope in \cite{Banerjee:2018tun}, is independent of associahedron embedding. 
As in the case of associahedron and $\phi^3$ amplitudes, the canonical form associated with the accordiohedron can be used to obtain n-particle planar scattering amplitude of the theory. However there is a key difference with the associahedron picture. Just like in the case of Stokes polytope and $\phi^4$ theory, the form associated with a single polytope only yields some of the channel-contributions in such a way that a weighted sum over the polytopes produces complete amplitude $\mathcal{M}_n$. We discuss these weights in section \ref{weights}. 

In section \ref{factorization}, we show that exactly as in the case of associahedron and $\phi^3$ theory and Stokes polytope and $\phi^4$ theory, factorization properties of accordiohedron imply the on-shell factorization of scattering amplitudes.

\section{Accordiohedron and mixed vertices}\label{accordiohedron}
\begin{figure}[H]
    \centering
    \includegraphics[scale=0.28]{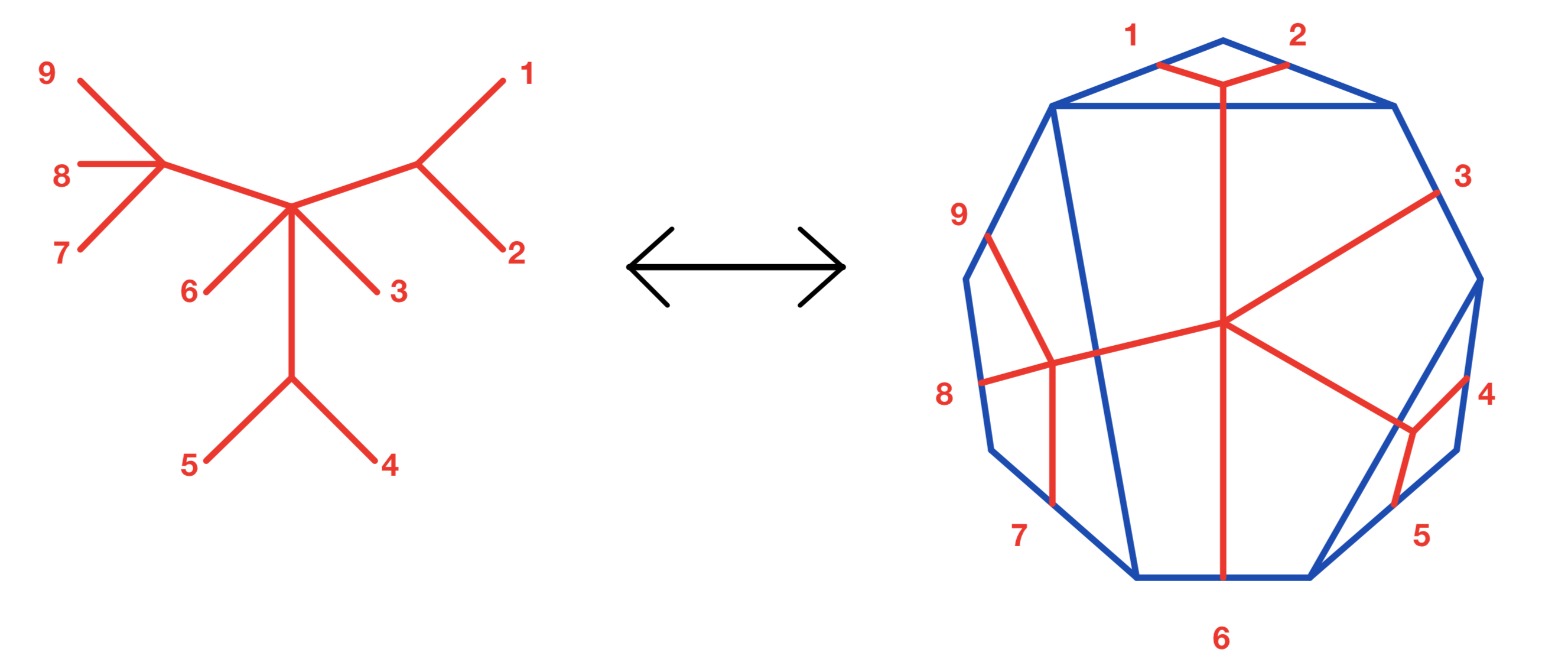}
    \caption{Correspondence between Feynman diagrams and dissections of polygon.}
    \label{fig1}
\end{figure}

In \cite{Arkani-Hamed:2017mur} and \cite{Banerjee:2018tun}, the Feynman diagrams in $\phi^3$ theory and in $\phi^4$ theory were associated with vertices of the simple polytopes, associahedron and Stokes polytope respectively. These associations were achieved by relating the Feynman diagrams in $\phi^3$ theory and in $\phi^4$ theory with triangulations and quadrangulations of polygons respectively. This relation can be extended to arbitrary vertices and combinations thereof. Recently, the extension to $\phi^p$ vertices was carried in \cite{Raman:2019utu}. We extend this to theories with mixed vertices. As with $\phi^3$, $\phi^4$ and $\phi^p$ theory, there is a one-to-one correspondence between tree-level $n$-point Feynman diagrams of a general planar scalar field theory and dissections of an $n$-gon into triangles, quadrilaterals and other $p$-gons. As an example, consider the Feynman diagram in figure \ref{fig1}, which has 2 cubic, 1 quartic and 1 quintic vertex, and consider the polygon on the right. Each side of the polygon corresponds to an external line, each diagonal of the dissection corresponds to a propagator, and each $p$-gon region of the dissection corresponds to a $p$-vertex.

Our goal is to locate a (convex realization of) simple polytope in kinematic space\footnote{We want a simple polytope because simple polytopes have a natural differential form associated with them. This natural form is our analogue of the scattering form in \cite{Arkani-Hamed:2017mur}}\footnote{A $d$-dimensional simple polytope is a polytope each of whose vertices are adjacent to exactly $d$ edges and $d$ co-dimension one facets} whose vertices are in one-to-one correspondence with all the dissections of an $n$-gon into triangles, quadrilaterals, and so on. The condition on the polytope to be simple is equivalent to requiring that the dimension of the polytope is the same as the number of propagators in a single channel. Just as in the case of Stokes polytope \cite{Banerjee:2018tun}, naively writing down all possible Feynman diagrams will give us more vertices than we need for a single simple polytope - a fact that is connected with the non-uniqueness of these polytopes in a given dimension. And just as in \cite{Banerjee:2018tun}, we have to define a notion of `compatibility' with a given Feynman diagram to rid some of the dissections to get a simple polytope. The definition of the accordiohedron, which we will give now, does precisely this.\\

\begin{figure}[H]
    \centering
    \includegraphics[scale=0.22]{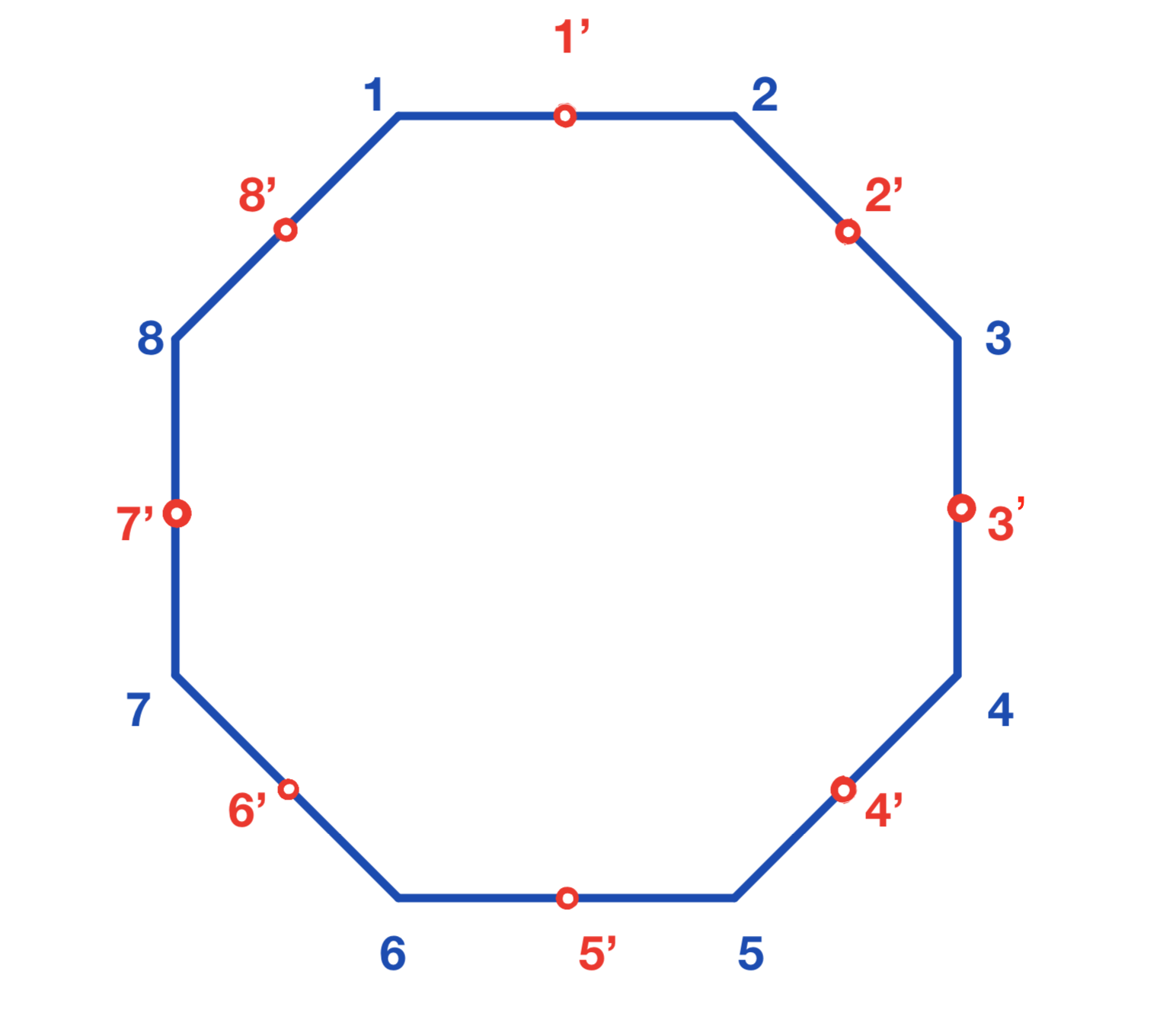}
    \caption{Dual polygon}
    \label{fig2}
\end{figure}

\begin{figure}[H]
    \centering
    \includegraphics[scale=0.25]{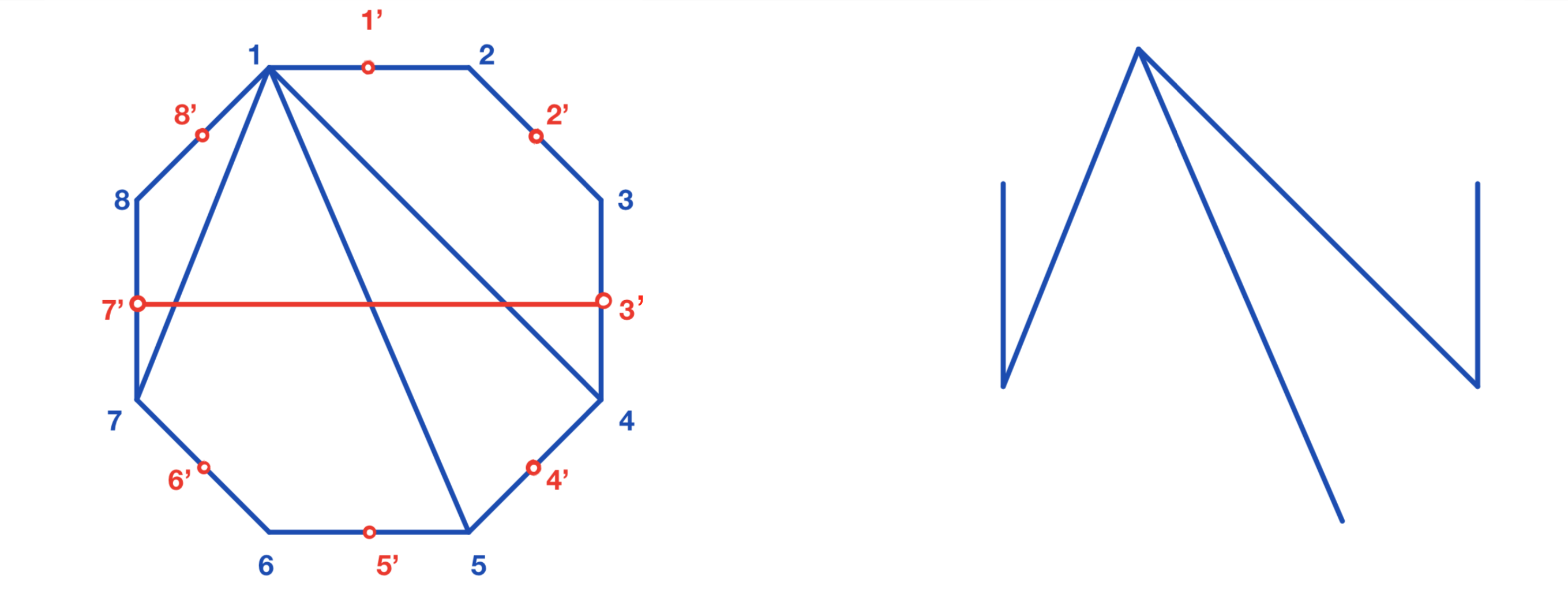}
    \caption{Reference dissection(in blue) and diagonal ($3',7'$) on the left and their cut on the right.}
    \label{fig3}
\end{figure}

Let $P$ be a convex polygon with $n$ vertices. We will call this polygon the solid polygon and label its vertices with $1,2,…,n$. Now we consider the polygon whose vertices are the mid-points of the sides of $P$. We will call this polygon the dual of $P$ or the hollow polygon. We label the midpoint of side $(i,i+1)$ by $i’$. (See figure \ref{fig2}).
A cut $C((i’,j’), D)$ of the hollow diagonal $(i’,j’)$ is a set comprising the sides $(i,i+1)$ and $(j,j+1)$ of the solid polygon along with the diagonals of dissection $D$ of the solid polygon which intersect the diagonal $(i’,j’)$. We say the hollow diagonal $(i’,j’)$ is compatible with the dissection $D$ if the cut $C((i’,j’), D)$ is connected (see figure \ref{fig3}). 
A $D$-accordion dissection is a dissection of the hollow polygon consisting of diagonals compatible with the dissection $D$. We define $D$-accordiohedron to be the simple polytope $\mathcal{AC}(D)$ whose vertices are all the $D$-accordion dissections \cite{manneville2019geometric}. 

If the dissection $D$ involves $d$ diagonals, then the accordiohedron $\mathcal{AC}(D)$ is a $d$-dimensional simple polytope. Its vertices are given by $D$-accordion dissections, each of these dissections consists of $d$ diagonals. Two vertices are adjacent to each other if they share all but one diagonal. This gives a one-dimensional boundary of $\mathcal{AC}(D)$ which is given by the shared $d-1$ diagonals. Similarly the two-dimensional boundaries of $\mathcal{AC}(D)$ are given by $ d-2$ diagonals and so on, finally each co-dimension one boundary facet corresponds to diagonals which are compatible with the reference dissection. 

Curiously, despite the seemingly abstract nature of this definition, the foregoing notion of compatibility makes contact with previously established details regarding the Stokes polytope and associahedron. More precisely, if the collection of considered dissections correspond to all possible quadrangulations, then the above notion of compatibility agrees with the definition of $Q$-compatibility (ref. \cite{Banerjee:2018tun}) and if the dissections are labelled by all possible complete triangulations of the polygon, the accordiohedron $\mathcal{AC}(T)$ for any triangulation $T$ is the associahedron\footnote{This latter point is not so difficult to see. Indeed, any triangulation saturates the polygon with dissections. One can immediately be convinced that any diagonal of the hollow polygon will supply a connected subgraph of the triangulation.}. 

To further understand the notion of compatibility and relate it to mutation of Feynman diagrams, we use the correspondence between dissections of $n$-gons and Feynman diagrams and give an equivalent definition of compatibility in terms of Feynman diagrams. The side $(i,i+1)$, which was labelled by $i'$ is associated with the external line with momentum $p_i$. The diagonal $(i,j)$ is associated with the propagator through which the momentum $(p_i+ p_{i+1}+ \cdots + p_{j-1}) = (p_j+ p_{j+1}+ \cdots + p_{i-1}) $ flows. In some sense, it is the propagator which connects the external lines $p_i$ and $p_j$. Thus the pair of external lines with momenta $p_i$ and $p_j$ is associated with the propagator through which the momentum $(p_i+ p_{i+1}+ \cdots + p_{j-1})$ flows. This correspondence is reflected in the one to one correspondence between the diagonals of the hollow polygon, and the diagonals of the solid polygon, the diagonal $(i',j')$ is related to the diagonal $(i,j)$. 

In a Feynman diagram, we say an external line or a propagator is adjacent to another external line or a propagator if 1) they are attached to the same vertex and 2) at that vertex they come one after the other if we go around the vertex in a cyclic order. For example, in figure \ref{fig1} the external line $p_7$ is adjacent to external line $p_8$ and the propagator carrying momentum $(p_7 + p_8 + p_9)$, but it is not adjacent to any other external line or propagator, while the external line $p_8$ is adjacent only to $p_7$ and $p_9$. Given a reference Feynman diagram, we say the external line $p_i$ is connected to the external line $p_j  (j\neq i \pm 1 \mod n )$ if we can get a sequence of propagators starting with a propagator adjacent to $p_i$ and terminating with a propagator adjacent to $p_j$ such that consecutive propagators are adjacent to each other. The hollow diagonal $(i',j')$ is compatible with the given Feynman diagram if the external line $p_i$ is connected to the external line $p_j$. For example, in figure \ref{fig1}, it is clear that (2,9) is a compatible diagonal and $p_2$ and $p_9$ are connected, while $p_8$ is not adjacent to any propagator and hence $(i,8)$ for any $i$ cannot be a compatible diagonal.  We can similarly use this language to understand the notions of $Q$-compatibility in \cite{Banerjee:2018tun} and mutation in \cite{Arkani-Hamed:2017mur}.

Using the correspondence between diagonals of the $n$-gon and the planar propagators we label the diagonal $(i,j)$ by the planar propagator $X_{ij}$ where, $$ X_{ij}= s_{i,i+1,\cdots j-1} = (p_i+ p_{i+1}+ \cdots + p_{j-1})^2.$$ From now on, unless otherwise stated we don't make a distinction between the diagonal $(i,j)$ and the propagator $X_{ij}$. Now we can define the canonical projective form associated with the accordiohedron in terms of these planar variables. 

In general we can define a projective form for any simple polytope\cite{Salvatori:2018aha}. The canonical form is given by $$ \Omega\left[\mathcal{P}\right]= \sum_{v\in \mathcal{P}} sign(v) \bigwedge_{ X \in v} \mathrm{d}\log(X),  $$ where $v$ is a vertex of the polytope $\mathcal{P}$.
The ordering of the facets $X$ and $sign(v)$ are fixed such that the form is projective. A simple rule to ensure projectivity is as follows. Suppose $v$ and $v'$ are adjacent vertices. Then they are given by intersections of two sets facets $A$ and $A'$ which have all but two elements same. Let's call these two elements $X$ and $X'$. Once the facets in the wedge product are ordered so that $X$ and $X'$ are in same position and the remaining elements which are there in both $A$ and $A'$ take same positions, we must have $sign(v)= -sign(v')$.
 
The $n$-point scattering amplitude of the theory with polynomial scalar interactions of the form $\sum_{i=3}^{N}\lambda_i \phi^i $ will have terms at order $ \lambda_{3}^{v_3} \cdots \lambda_{N}^{v_N} $, for all $v_i$s that satisfy $(n - 2) = \sum_{i=3}^N (i-2)v_i .$ The accordiohedra associated with dissections of $n$-gon into $a_3$ triangles, $a_4$ quadrilaterals  and so on, will contribute to the term at the order $\lambda_{3}^{a_3} \cdots \lambda_{N}^{a_N}$. To get the full scattering amplitude, we will have to sum over all possible dissections. For example, to get the full six-point scattering amplitude we will have to sum over contributions coming from associahedra associated with dissections of hexagon into four triangles, two triangles and a square, two squares, a pentagon and a triangle, and one hexagon. Here we compute the contribution coming from dissections of hexagon into two triangles and a square.

Before moving on to calculations that will help us examine this idea more carefully, a couple of points of importance should be noted. Predominantly, the accordiohedron, like the Stokes polytope is not uniquely determined by the dimension. In addition to these facts, the accordiohedron is defined by conditions that do not depend on the nature of the dissections. Naturally, one is led to suspect that it may indeed be the correct positive geometry for general scalar field theories at tree level. We are left then, with a task very similar to what was encountered by the authors of ref. \cite{Banerjee:2018tun} and \cite{Raman:2019utu}, in the case of Stokes polytopes and accordiohedra, {\it viz.}, the various accordiohedra of a given dimension have to be accordingly weighted to ensure that the full amplitude is obtained.

 We will proceed to look at a simple example in order to make the definition of the accordiohedron less impenetrable to the reader. Consider  five particle scattering process at tree level. We specialize to the case when the amplitude is generated by one cubic and one quartic vertex. To be more precise, we consider the interaction vertices,

\begin{equation}
    \mathcal{L}_{int} = \frac{1}{3!}\lambda_3\phi^3 + \frac{1}{4!}\lambda_4 \phi^4.
\end{equation}

Now for the five particle amplitude with one cubic and one quartic contribution, the amplitude will be dressed by $\lambda_3\lambda_4$. The amplitude is readily written down,

\begin{equation}
    \mathcal{M}^{5}_{\lambda_3\lambda_4}(p_1,...,p_5) = \lambda_3\lambda_4 \left(\frac{1}{s_{12}} + \frac{1}{s_{23}} + \frac{1}{s_{34}} + \frac{1}{s_{45}} + \frac{1}{s_{51}}\right).
\end{equation}

To avoid a profusion of notation, the coupling constants will be dropped henceforth, with the understanding that they are suitably represented on the symbol for the amplitude. 

Now, let us look at a pentagon with the dissection $(13)$. This corresponds to the channel giving the first term in the foregoing equation. Now, for this simple case, the only two compatible dissections of this form are $(1',3')$ and $(2',5')$. The reader will recognize here that the polytope so obtained is just a line, which coincides with the associahedron and Stokes polytope of the same dimension, courtesy of the fact that it is $1$ dimensional. We will illustrate now the rudiments of how the scattering form and embedding are done, with this simple illustration hopefully giving the reader the basic idea of how the rest of the paper's calculations are structured. 

The planar scattering form is developed by considering,

\begin{equation}
    \Omega\left[\mathcal{AC}(13)\right] = d\log(X_{13}) \pm d\log(X_{25}).
\end{equation}
The sign must be chosen such that the form $\Omega\left[\mathcal{AC}(13)\right] $ is projective (invariant under $X_{ij} \rightarrow \alpha(X)X_{ij} $), and hence we have,

\begin{equation}
    \Omega\left[\mathcal{AC}(13)\right] = d\log(X_{13}) - d\log(X_{25}).
\end{equation}

Now how is this accordiohedron to be embedded in kinematical space in order to supply the amplitude? We expect that the facets of the accordiohedron go to zero on the boundary of the embedded accordiohedron. We already have an embedded polytope on whose boundaries the facets of accordiohedron go to zero, the kinematic associahedron. So first, we put the constraints corresponding to the kinematic associahedron \cite{Arkani-Hamed:2017mur}, namely,

\begin{equation}
    s_{ij} = -c_{ij},
\end{equation}

where $(ij)$ belongs to $\lbrace{(13), (14), (24)\rbrace}$. Expanded in terms of the planar variables $X_{ij}$ these are,

\begin{equation}
    X_{13} + X_{24} - X_{14} = c_{13}, \label{ass_con1}
\end{equation}

\begin{equation}
    X_{14} + X_{25} - X_{24} = c_{14} \label{ass_con2}
\end{equation}

and 

\begin{equation}
    X_{24} + X_{35} - X_{25} = c_{24}. \label{ass_con3}
\end{equation}

The first two of these added together, give,

\begin{equation}
    X_{13} + X_{25} = c_{13} + c_{14}.
\end{equation}

Which gives, without much more work, the amplitude due to the consequence $dX_{13} = -dX_{25}$. The three equations above describe however a two dimensional polytope. One of three constraints, namely $X_{35} =d_{35}$, $X_{24} = d_{24}$ and $X_{14} = d_{14}$ may be set to recover the accordiohedron. Note that, while constraints \eqref{ass_con1} - \eqref{ass_con3} are the same for any reference dissection, the additional constraint depends on the dissection and should not be arbitrarily chosen. This way of writing constraints is rather ad-hoc, and we provide a better way in section \ref{embedding}.

The remaining accordiohedra can be obtained by permuting $(13)$. In doing so, each channel is seen to contribute twice. We set the weights so that the residue in each channel is one. Consequently, the weights in this case are uniquely fixed as $\frac{1}{2}$,

\begin{equation}
    \mathcal{M}^{5}_{\lambda_3\lambda_4}(p_1,...,p_5) = \frac{1}{2}\mathcal{M}_{(13)} + \frac{1}{2}\mathcal{M}_{(24)} +...+\frac{1}{2}\mathcal{M}_{(52)}.
\end{equation}

In the next section, we present a more nontrivial $6$ point amplitude that has two cubic and one quartic vertex.

\section{Accordiohedra for a 6 point amplitude}\label{sixpoint}
\begin{figure}[H]
    \centering
    \includegraphics[scale=0.25]{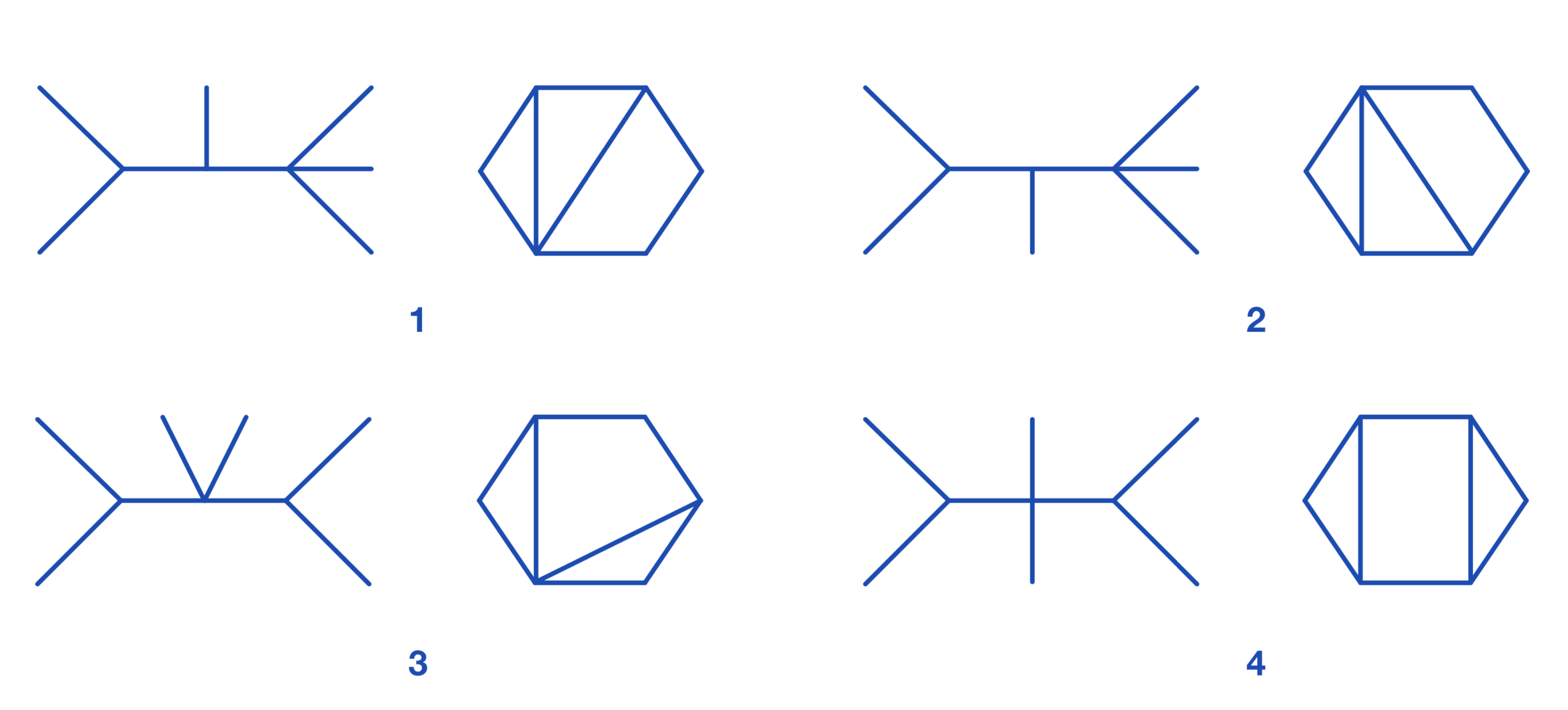}
    \caption{Topologically in-equivalent Feynman diagrams and corresponding dissections}
    \label{6ineqfeyn}
\end{figure}

\begin{figure}[H]
    \centering
    \includegraphics[scale=0.45]{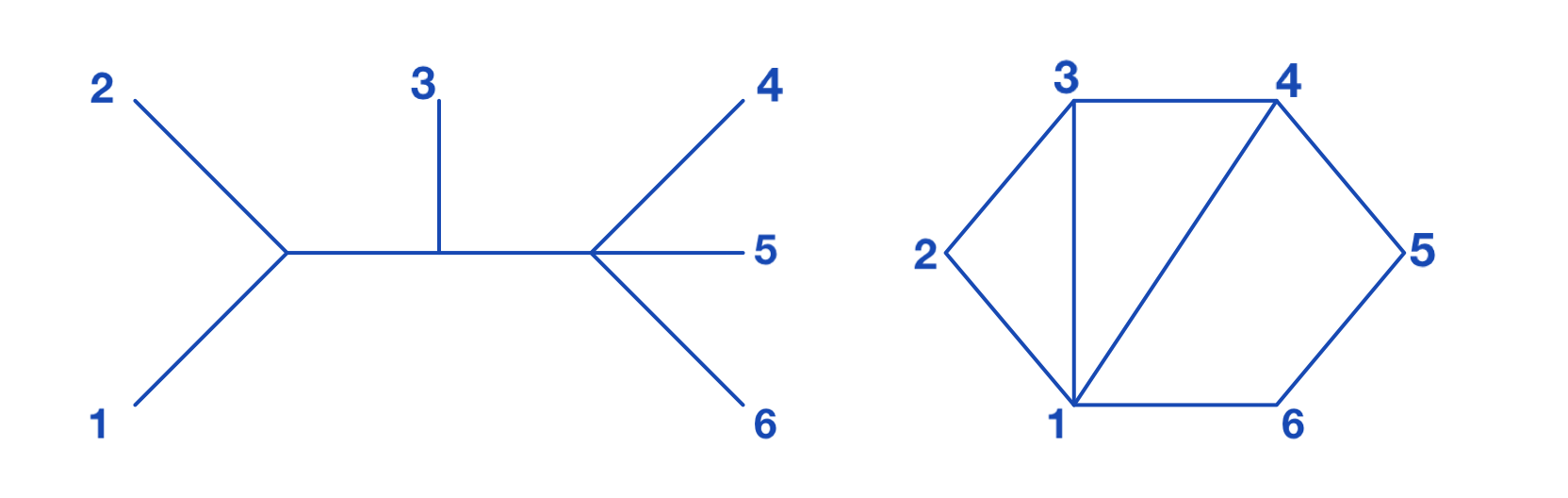}
    \caption{Dissection for the given ordering}
    \label{fig5}
\end{figure}

\begin{figure}[H]
    \centering
    \includegraphics[scale=0.25]{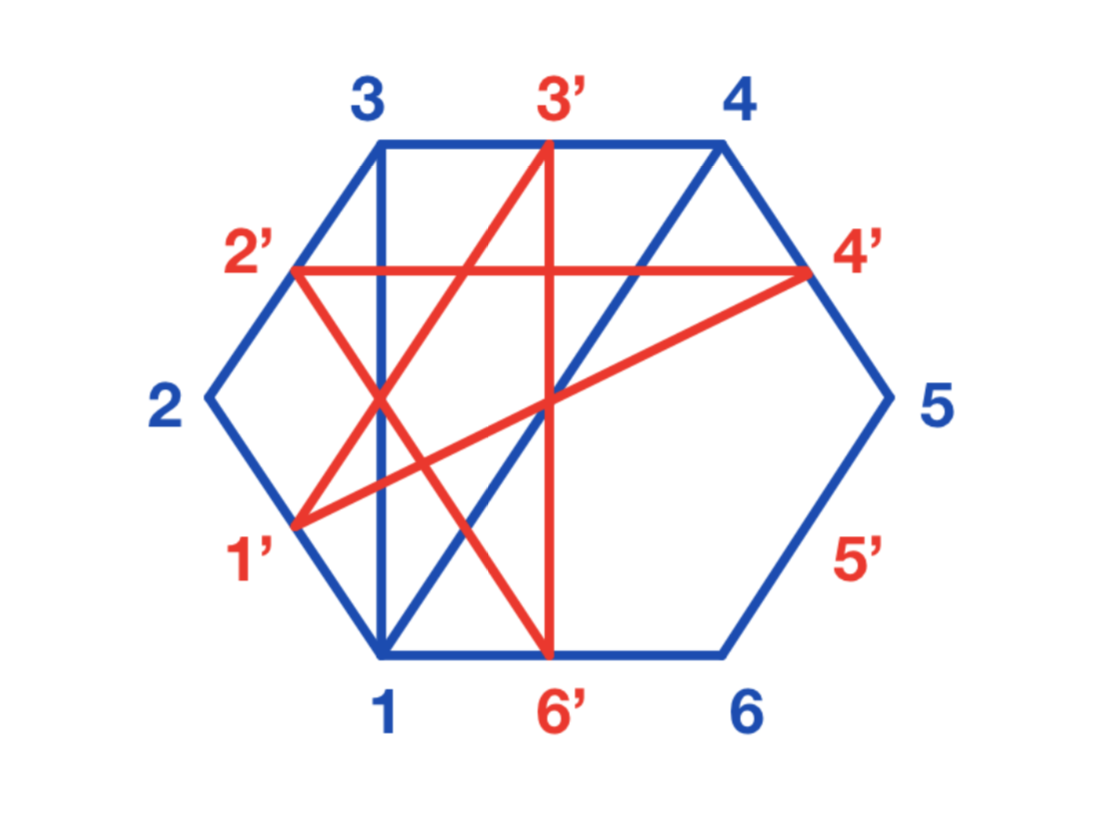}
    \caption{Diagonals compatible with $((1,3)(1,4))$}
    \label{fig6}
\end{figure}

\begin{figure}[H]
    \centering
    \includegraphics[scale=0.35]{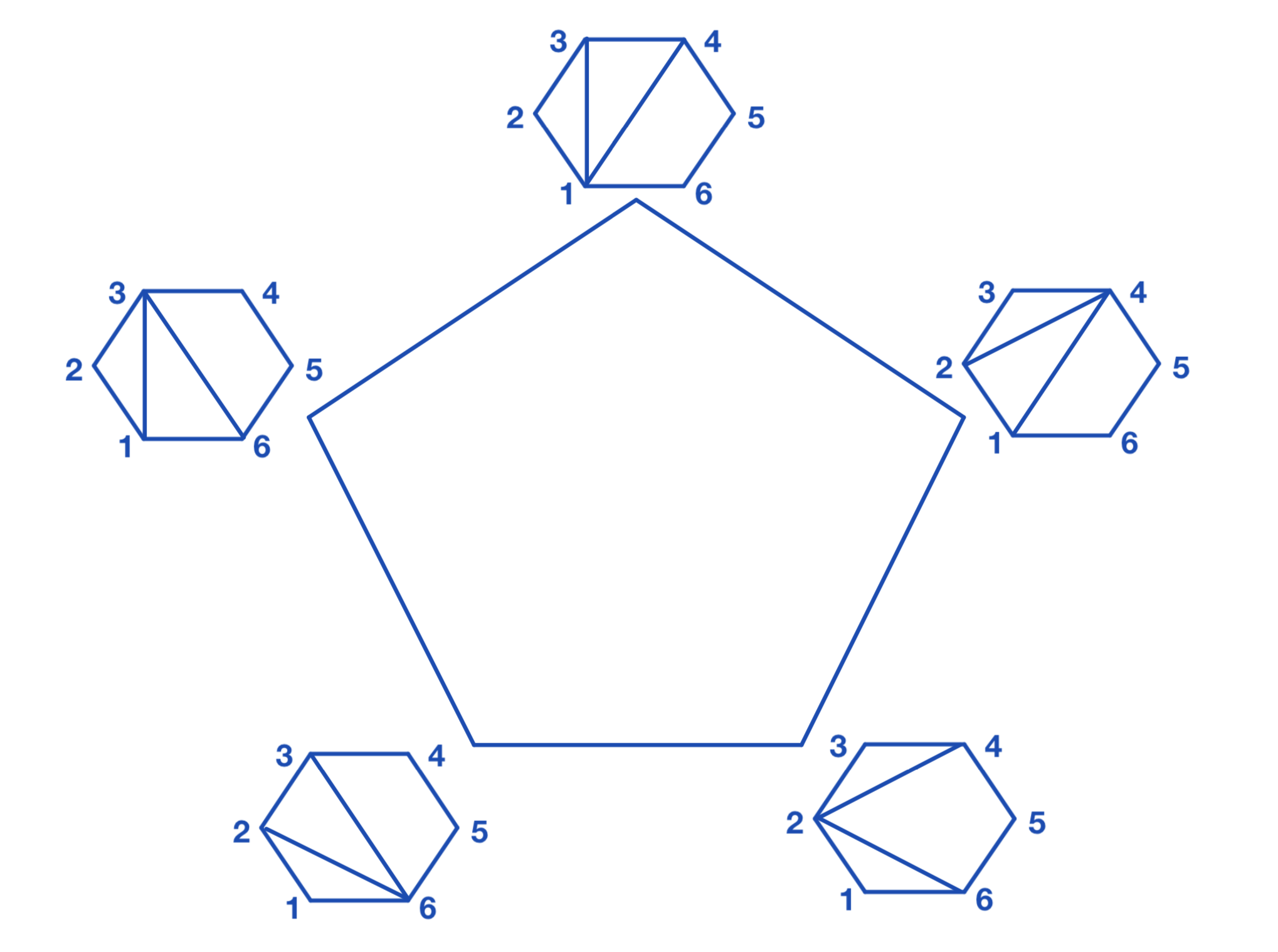}
    \caption{Accordiohedron for the dissection $((1,3)(1,4))$}
    \label{Accordion1314}
\end{figure}

Now, let's look at accordiohedra for 6 point amplitude with one four point vertex and two three point vertices . There are four topologically inequivalent Feynman diagrams (see figure \ref{6ineqfeyn}). Correspondingly there are four topologically inequivalent dissections of a hexagon into two triangles and a quadrilateral\footnote{We say two dissections are topologically equivalent if one is obtained form the other by cyclically permuting the vertices of the polygon.}. 
We will start by finding the accordiohedron for the first diagram. We will first look at a particular ordering given in the figure \ref{fig5}. Accordiohedra for cyclically permuted diagrams can be obtained by similarly permuting dissections which make up the vertices of the accordiohedron for the ordered diagram.
The diagonals compatible with the given dissection are $ (1',3')$, $(1',4')$, $(2',4')$, $(2',6')$ and $(3',6')$ as shown in figure \ref{fig6}. 
\noindent Hence the vertices of the accordiohedron are $((1,3)(1,4))$, $((2,4)(1,4))$, $((2,4)(2,6))$, $((2,6)(3,6))$ and $((1,3)(3,6)) $.
We start with the dissection $((1,3)(1,4))$ and get other compatible dissections by flipping the diagonals one by one. Given any dissection, a \textit{flip} is defined as replacing one of the diagonals of the dissection with a non-crossing, compatible diagonal to get another compatible dissection. This will tell us which vertices are adjacent to each other. By flipping $(1,3)$ to $(2,4)$ we get $((2,4)(1,4))$ similarly by flipping $(1,4)$ to $(3,6)$ we get $((1,3)(3,6))$. Further flipping $(1,4)$ in $((2,4)(1,4))$ and $(1,3)$ in $((1,3)(3,6))$ we get $((2,4)(2,6))$ and $((2,6)(3,6))$ respectively. Thus the geometric realization of this accordiohedron is given in figure \ref{Accordion1314}.

Using similar procedures, we can get the accordiohedron for the remaining Feynman diagrams. For the sake of brevity, we simply note the accordiohedra for the remaining reference dissections:
Accordiohedra for the second diagram are $$ \mathcal{AC}((1,3)(3,6)) = \left\{ ((1,3)(3,6)),((2,6)(3,6)),((2,6)(2,5)),((2,5)(1,5)),((1,3)(1,5))  \right\}$$ and its cyclic permutations. 
Accordiohedra for the third diagram are $$ \mathcal{AC}((1,3)(1,5)) = \left\{ ((1,3)(1,5)),((2,5)(1,5)),((2,5)(2,6)),((2,6)(4,6)),((1,3)(4,6))  \right\}$$ and its cyclic permutations.
Accordiohedra for the fourth diagram are $$ \mathcal{AC}((1,3)(4,6)) = \left\{ ((1,3)(4,6)),((2,6)(4,6)),((2,6)(3,5)),((1,3)(3,5))  \right\}$$ and its cyclic permutations.

Now we proceed to computing the canonical form associated with the accordiohedron. As mentioned in the previous section, we have a projective form (up to overall sign) $\Omega(\mathcal{P})$, which is a sum of forms for any simple polytope $\mathcal{P}$. In the case of the accordiohedron, we use the \textit{flipping} operation to fix the relative signs among the sum of forms. So we fix the order of forms on a given vertex, and preserving this order any flip will come with a factor of $(-1)$. $n$ such flips will give a factor corresponding to $(-1)^n$. This is a natural generalisation of mutation in the case of the associahedron and the Stokes polytope.

Going back to the example in figure \ref{Accordion1314}, the canonical form associated with this accordiohedron is,
\begin{equation}
\begin{split}
    \Omega\big[\mathcal{AC}\big((1,3),(1,4)\big)\big] = {} & d \ln X_{13}\wedge d \ln X_{14} - d \ln X_{24}\wedge d \ln X_{14} - d \ln X_{13}\wedge d \ln X_{36} \\&+ d \ln X_{24}\wedge d \ln X_{26} + d \ln X_{26}\wedge d \ln X_{36}.
\end{split}
\end{equation}

Similarly, the canonical forms associated with the remaining accordiohedra in the six particle case may be written down as follows,

\begin{equation}
\begin{split}
    \Omega\big[\mathcal{AC}\big((1,3),(3,6)\big)\big] = {} & d \ln X_{13}\wedge d \ln X_{36} - d \ln X_{13}\wedge d \ln X_{15} - d \ln X_{26}\wedge d \ln X_{36}\\& + d \ln X_{25}\wedge d \ln X_{15} + d \ln X_{26}\wedge d \ln X_{25},
\end{split}
\end{equation}

\begin{equation}
\begin{split}
    \Omega\big[\mathcal{AC}\big((1,3),(1,5)\big)\big] = {} & d \ln X_{13}\wedge d \ln X_{15} - d \ln X_{13}\wedge d \ln X_{46} - d \ln X_{25}\wedge d \ln X_{15}\\& + d \ln X_{25}\wedge d \ln X_{26} + d \ln X_{26}\wedge d \ln X_{46},
\end{split}
\end{equation}
and
\begin{equation}
\begin{split}
    \Omega\big[\mathcal{AC}\big((1,3),(4,6)\big)\big] = {} & d \ln X_{13}\wedge d \ln X_{46} - d \ln X_{26}\wedge d \ln X_{46} + d \ln X_{26}\wedge d \ln X_{35}\\& - d \ln X_{13}\wedge d \ln X_{35}. 
\end{split}
\end{equation}

Now that we have the canonical forms for all the accordiohedra, we have to embed these accordiohedra in kinematic space and pullback these canonical forms onto the kinematic accordiohedra. To then get the scattering amplitude, we need to weight each form and add them, as shown in section \ref{weights}. In the following section, we offer a slightly different way (or a more general way that reduces to associahedron constraints) of looking at the constraints (unlike in section \ref{accordiohedron}) that embed the accordiohedra and postpone the calculation of finding the amplitude for this example until the end of the next section.

\section{Embedding accordiohedron in kinematic space}\label{embedding}

In section \ref{accordiohedron}, we gave the accordiohedron as the positive geometry for general planar scalar field theories. Just like associahedron and Stokes polytope, we want to embed accordiohedron in the kinematic space. In \cite{Banerjee:2018tun}  the Stokes polytope was embedded by first considering the associahedron constraints and then imposing some additional constraints. The additional constraints were given by putting some of the propagators which do not occur in $\phi^4$ theory to constant. This rule is rather ad-hoc for theories with combinations of cubic and higher interaction vertices. 

We also note that as higher order scalar interactions are independent of the cubic interactions, it is desirable to have a convex realization of the accordiohedron in (the positive region of) kinematic space which does not rely on existence of associahedron. We would thus like to have embeddings of positive geometries of theory with combinations of cubic, quartic and higher interaction vertices to be independent of the embedding of positive geometries of theory with cubic vertices. In this section, we generalize the convex realization of associahedron given in the works of Loday \cite{loday2004realization} and Nima Arkani-Hamed, et al. \cite{Arkani-Hamed:2017mur}, and provide an embedding of accordiohedron which is independent of associahedron constraints. In this paper, we illustrate our proposal with two examples with mixed vertices, two propagators and six external lines. More details of the embedding of the accordiohedron for a higher number of propagators and external lines will be given in an upcoming work \cite{Aneesh:2019cvt}. For a precise mathematical description of geometric realisation of accordiohedron, we refer to section 2 of \cite{padrol2019associahedra}.

In \cite{Arkani-Hamed:2017mur}, the authors first defined a region $\Delta$ by restricting the planar variables to the positive region. Then they intersected it with a region defined by setting the Mandelstam variables $s_{ij}$ which never occur in planar amplitudes, to constant. Here we give a novel way of looking at the constraints in the second step and generalize it to give an embedding of the accordiohedron in kinematic space.

A vertex of associahedron corresponds to all propagators of a Feynman diagram going on-shell, a one-dimensional facet (edge) of associahedron corresponds to all but one propagator of a Feynman diagram going on-shell, and so on. Finally, a co-dimension one facet of the associahedron corresponds to a propagator going on-shell. We embed the associahedron in the kinematic space such that the vertices of the embedded polytope are points in the kinematic space where all the propagators of the corresponding Feynman diagram go on-shell. A co-dimension one boundary of this embedded polytope is a set of points in kinematic space where the corresponding propagator goes on-shell. Two co-dimension one facets of associahedron are adjacent to each other only if there is a Feynman diagram which contains both the propagators associated with these facets \footnote{We say two co-dimension one facets are adjacent if they share a co-dimension two boundary.}. We will use these facts to understand the constraints of the kinematic associahedron.

The facet $X_{i,j+1}$ is adjacent to the facets $X_{i,k+1}$ and $X_{k,j+1}$ but the intersection of these three facets is empty. In other words, in planar $\phi^3$ theory we can have a Feynman diagram with propagators $X_{i,j+1}$ and $X_{i,k+1}$ or with propagators $X_{i,j+1}$ and $X_{k,j+1}$ but we can not have a Feynman diagram with propagators $X_{i,j+1}$, $X_{i,k+1}$ and $X_{k,j+1}$. To ensure this we put the constraints,
\begin{equation}\label{ascons1}
    X_{i,j+1}= X_{i,k+1}+ X_{k,j+1} - \epsilon_{i,j,k}, \hspace{0.7cm} \epsilon_{i,j,k} > 0,\hspace{0.15cm} 1 \leq i < k <j < n, \hspace{0.15cm} j-i < n-2 .
\end{equation}

These constraints ensure that the propagators $X_{i,j+1}$, $X_{i,k+1}$ and $X_{k,j+1}$ can never simultaneously be zero on associahedron.

Similarly, the two facets $X_{1,i+1}$ and $X_{i,n}$ can not be adjacent to each other. This is ensured by enforcing the constraints,
\begin{equation}\label{ascons2}
    X_{1,i+1} + X_{i,n} = \epsilon_{i} , \hspace{0.7cm}  \epsilon_{i}>0,\hspace{0.15cm} 1<i<n-1.
\end{equation}
To enforce constraints \eqref{ascons1} and \eqref{ascons2} it is enough to enforce the following constraints,
\begin{equation}\label{ascons}
    X_{i,j+1} = X_{1,j+1} - X_{1,i+1} + \epsilon_{i,j}, \hspace{0.7cm}  \epsilon_{i,j}>0,\hspace{0.15cm} 1<i<j<n.  \footnote{We can think of \eqref{ascons2} as defining a cuboid in a $(n-3)$ dimensional space. The other $\frac{(n-3)(n-4)}{2}$ constraints define planes which truncate this cuboid. This gives us the following properties of the Loday associahedron\cite{ceballos2015many}; The geometric realization of the associahedron has $(n-3)$ pairs of parallel facets and in a basis defined by these $(n-3)$ pairs, the normal vectors to the truncating planes have coordinates in $\{0,\pm1 \}$.}
\end{equation}
Now we will show that these $\frac{(n-2)(n-3)}{2}$ constraints are equivalent to the associahedron constraints given in \cite{Arkani-Hamed:2017mur}, $ s_{i,j} = - c_{i,j} $ for $c_{i,j}>0$ and $ 1 \leq i < j < n$.  Let us consider the sum $a_{ij}(k) = \sum_{l=i+1}^j s_{kl} $. Using $s_{kl} =   X_{k+1,l}  - X_{k+1,l+1} - X_{k,l} + X_{k,l+1}  $ we make this sum into a telescopic sum to get $a_{ij}(k) = X_{k+1,i+1} - X_{k+1,j+1} - X_{k,i+1} + X_{k,j+1}  $. Now we sum $a_{ij}(k)$ from $k=1$ to $k= i-1$ to get 
\begin{equation}\label{id1}
     \sum_{k= 1}^{i-1} \sum_{l=i+1}^j s_{kl} = X_{i,i+1}- X_{1,i+1} + X_{1,j+1} - X_{i,j+1} = -(X_{i,j+1} - (X_{1,j+1} - X_{1,i+1}) ).
\end{equation}
Using \eqref{id1} it is easy to see that if we set $ \sum_{k= 1}^{i-1} \sum_{l=i+1}^j s_{kl}= -\sum_{k= 1}^{i-1} \sum_{l=i+1}^j c_{kl} = -\epsilon_{i,j} $ we get our constraints \eqref{ascons}.

Now we will use similar logic to embed accordiohedron associated with a Feynman diagram in the kinematic space. 
We first restrict the planar variables to the positive region by setting,
\begin{equation}
      X_{i,j} \geq 0 \hspace{0.2cm} \text{ for all } 1\leq i <j \leq n. 
\end{equation}
If the Feynman diagram has quartic or higher order vertices, then some propagators don't occur as a facet of accordiohedron associated with it. We simply put such planar variables to a positive constant. For the remaining propagators, we use the above logic. Let's look at two dissections of the example in section \ref{sixpoint}:
\subsection*{Dissection 1}

Consider the six-point Feynman diagram with propagators $X_{1,3}$, $X_{1,4}$. The propagators $X_{1,3}$, $X_{1,4}$, $X_{2,4}$, $X_{2,6}$ and $X_{3,6}$ are compatible the six-point Feynman diagram $X_{1,3}$, $X_{1,4}$. We put all other planar variables to constant. In a planar theory, we can have a Feynman diagram with propagators $X_{1,3}$ and $X_{1,4}$ or with propagators $X_{2,4}$ and $X_{1,4}$ but, we can not have propagators $X_{1,3}$, $X_{1,4}$ and $X_{2,4}$ together in a single Feynman diagram. Therefore we enforce the constraint $X_{1,3} + X_{2,4} = X_{1,4} + c_1 $. We can not have a six-point Feynman diagram with propagators $X_{1,3}$ and $X_{2,6}$ or a Feynman diagram with propagators $X_{1,4}$ and $X_{3,6}$. Therefore we enforce the constraints $X_{1,3}+ X_{2,6} = c_2$ and $ X_{1,4} + X_{3,6} = c_3$. \footnote{We must have $c_2>c_1$ and $c_1,c_2,c_3>0$.} The pullback of the canonical form onto the embedded kinematic accordiohedron is then,
\begin{equation}
\begin{split}
F^*\Omega\big[\mathcal{AC}_k\big((1,3),(1,4)\big)\big] = {} & \left( \frac{1}{X_{13} X_{14}} + \frac{1}{X_{24} X_{14}} + \frac{1}{X_{13} X_{36}} + \frac{1}{X_{24} X_{26}} +\frac{1}{X_{26} X_{36}} \right) \\ & \times d X_{13} \wedge d X_{14} \ .
\end{split}
\end{equation}

\subsection*{Dissection 4}
Similarly, let us look at the same diagram, but with propagators $X_{13}$, $X_{46}$, which corresponds to the fourth diagram in figure \ref{6ineqfeyn}. The other compatible propagators are $X_{26}$ and $X_{35}$. The rest is set to constant. Then, applying the same logic as before, we have the following equations : $X_{46} = - X_{35} + d_1$ and $X_{26} = - X_{13} + d_2$. The pullback of the canonical form onto the embedded kinematic accordiohedron is then,
\begin{equation}
\begin{split}
F^*\Omega\big[\mathcal{AC}_k\big((1,3),(4,6)\big)\big] = {} & \left( \frac{1}{X_{13} X_{46}} + \frac{1}{X_{26} X_{46}} + \frac{1}{X_{26} X_{35}}  + \frac{1}{X_{13} X_{35}} \right) d X_{13} \wedge d X_{46} \ .
\end{split}
\end{equation}

Similarly, for the other two dissections, we directly write the pullback of the forms,
\begin{equation}
\begin{split}
F^*\Omega\big[\mathcal{AC}_k\big((1,3),(3,6)\big)\big] = {} & \left( \frac{1}{X_{13} X_{36}} + \frac{1}{X_{13} X_{15}} + \frac{1}{X_{26} X_{36}} + \frac{1}{X_{25} X_{15}} +\frac{1}{X_{26} X_{25}} \right) \\ & \times d X_{13} \wedge d X_{36} \ ,
\end{split}
\end{equation}
and 
\begin{equation}
\begin{split}
F^*\Omega\big[\mathcal{AC}_k\big((1,3),(1,5)\big)\big] = {} & \left( \frac{1}{X_{13} X_{15}} + \frac{1}{X_{13} X_{46}} + \frac{1}{X_{25} X_{15}} + \frac{1}{X_{25} X_{26}} +\frac{1}{X_{26} X_{46}} \right) \\ & \times d X_{13} \wedge d X_{15} \ .
\end{split}
\end{equation}

\section{Weights of Feynman diagrams}\label{weights}
Now that we have the embedding of accordiohedron inside the kinematical space, we pull back the canonical form of the accordiohedron defined in the kinematical space onto the embedded accordiohedron and get 
\begin{equation}
    F^*\Omega(\mathcal{AC}_k(D)) =  \omega(\mathcal{AC}_k(D)) \bigwedge_{(i,j)\in D} \mathrm{d}X_{ij}.
\end{equation}{}
We call the function $\omega(\mathcal{AC}_k(D)) $ the canonical function associated with the dissection D (or the corresponding Feynman diagram). The full amplitude is given by weighted sum of these canonical functions. $$ \widetilde{\mathcal{M}_n} = \sum_{D} \alpha_D \omega(\mathcal{AC}_k(D)).  $$
We fix the weights by demanding that all poles of $\widetilde{\mathcal{M}_n}$ come with residue one. This condition is equivalent to the following system of linear equations in $\alpha$s, 
\begin{equation}\label{weights1}
 \sum_{D} \alpha_D \delta_D(D_i) = 1 \hspace{1.5cm} \text{for all dissections $D_i$}.
\end{equation}
Where $\delta_D(D_i)$ tells you whether the dissection $D_i$ is in the accordiohedron associated with the dissection $D$ or not. That is, $\delta_D(D_i) = 1$ if $D_i \in \mathcal{AC}(D) $ and zero otherwise. In this system of linear equations, both the number of equations and the number of variables is equal to the number of dissections. Therefore, generically, we have at least one solution. To simplify the task of finding weights, we put one more condition on $\alpha$s, we demand that all topologically equivalent dissections have equal weights. This condition is based on the intuition that $\alpha_D$ only depends on the intrinsic (combinatorial) property of $\mathcal{AC}(D)$.

Since the weights $\alpha$ are same for topologically equivalent diagrams we can club all the topologically equivalent dissections in \eqref{weights1} and write \eqref{weights1} as sum over equivalence classes, 
\begin{equation}\label{weights2}
    \sum_{[D]} \alpha_{D} N_{[D]}(D_i) = 1 \hspace{1.5cm} \text{for all dissections $D_i$}.
\end{equation}
Where $[D]$ denotes the equivalence class of $D$ and $N_{[D]}(D_i)$ is the number of times the dissection occurs in accordiohedra of all dissections which are topologically equivalent to $D$, that is $N_{[D]}(D_i) = \sum_{D_j \in [D]} \delta_{D_j}(D_i) $. 
Now the number of equations is equal to the number of dissections and number of variables ($\alpha_D$) is equal to the number of equivalence classes, so it may seem that the conditions on weights are over constraining and we may not have a set of weights satisfying the above requirements. But we will argue that there is at least one solution to the above system of linear equations. We will argue that $N_{[D]}(D_i) = N_{[D]}(D_j) $ if $D_i$ and $D_j$ are topologically equivalent. 

It is easy to see that a dissection $D_i$ is in the accordiohedron $\mathcal{AC}(D)$ if and only if the cyclically permuted (topologically equivalent) dissection $D_j= \sigma \cdot D_i$ is in the accordiohedron $\mathcal{AC}(D')$ of the similarly cyclically permuted dissection $D'= \sigma \cdot D$. Hence 
\begin{equation} \label{weights3}
    \delta_{D}(D_i)=\delta_{\sigma \cdot D}(\sigma \cdot D_i) = \delta_{D'}(D_j).
\end{equation}{} 
Using \eqref{weights3} and the definition of $N_{[D]}(D_i)$ we get 
\begin{equation}\label{weights4}
    N_{[D]}(D_i) = \sum_{D_k \in [D]} \delta_{D_k}(D_i) = \sum_{D_k \in [D]} \delta_{\sigma \cdot D_k}(\sigma \cdot D_i) = \sum_{D_k \in [D]} \delta_{D_k}(\sigma \cdot D_i) =   N_{[D]}( D_j).
\end{equation}
Therefore if $D_i$ and $D_j$ are topologically equivalent, the equations associated with them are equivalent. Thus the number of in-equivalent equations is equal to the number of equivalence classes. Hence we have the same number of equations as variables.

We denote the number of times dissection equivalent to $D_i$ occur in the accordiohedron of $D$ by $M_{[D_i]}(D)$, that is $M_{[D_i]}(D) = \sum_{D_k \in [D_i]} \delta_D(D_k) $. We claim that 
\begin{equation}\label{weights5}
   \vert \text{Stab}_{C_n}(D)\vert N_{[D]}(D_i) = \vert \text{Stab}_{C_n}(D_i)\vert  M_{[D_i]}(D).
\end{equation}{}
Where $C_n$ is the cyclic group of order $n$ and Stab$_{C_n} (D)$ is the stabilizer of the $C_n$ action on the dissection $D$.
\begin{align*}
     \vert \text{Stab}_{C_n}(D)\vert N_{[D]}(D_i) &= \vert \text{Stab}_{C_n}(D)\vert \sum_{D_j \in [D]} \delta_{D_j}(D_i)   \\ 
     &=  \sum_{ \sigma \in C_n} \delta_{\sigma \cdot D}(D_i)   \\ 
     &=  \sum_{ \sigma \in C_n} \delta_{D}(\sigma ^{-1} \cdot D_i)   \\ 
    \vert \text{Stab}_{C_n}(D)\vert N_{[D]}(D_i)  &= \vert \text{Stab}_{C_n}(D_i)\vert \sum_{D_k \in [D_i]} \delta_D(D_k)
\end{align*}
Which proves our claim. Using \eqref{weights5} we can get the coefficients $N_{[D]}(D_i) $ in the system if linear equations \eqref{weights2} just by considering associahedra of the representatives of the equivalence classes. 
Now we will use the results we just derived to compute weights for 6 point amplitude with two cubic vertices and a quartic vertex. As discussed in section 3 there are four topologically in-equivalent dissections labelled by 1,2,3 and 4 in figure \ref{6ineqfeyn}. The size of stabilizer groups are $\vert \text{Stab}_{C_n}(D_1)\vert = 1 $, $\vert \text{Stab}_{C_n}(D_2)\vert =1 $, $\vert \text{Stab}_{C_n}(D_3)\vert = 1$ and $\vert \text{Stab}_{C_n}(D_4)\vert = 2$. In the accordiohedron associated with $D_1$ dissections equivalent to $D_1$ occur twice, dissections equivalent to $D_2$ occur twice, dissections equivalent to $D_3$ occur once and the dissections equivalent to $D_4 $ don't occur therefore the equation associated with the equivalence class of dissection $D_1$ is  
\begin{equation}
    2\alpha_{D_1} + 2\alpha_{D_2} + \alpha_{D_3} + 0\alpha_{D_4} = 1. 
\end{equation}{}
Similarly we can get following equations for the remaining equivalence classes,
\begin{equation}
    2\alpha_{D_1} + 2\alpha_{D_2} + \alpha_{D_3} + 0\alpha_{D_4} = 1, 
\end{equation}{} 
\begin{equation}
    \alpha_{D_1} + \alpha_{D_2} + 2 \alpha_{D_3} + 1\alpha_{D_4} = 1,
\end{equation}{}
\begin{equation}
    0\alpha_{D_1} + 0\alpha_{D_2} + 2\alpha_{D_3} + 2\alpha_{D_4} = 1. 
\end{equation}{}
The solution of this system of linear equations is $\alpha_{D_1} = \frac{1}{2}-a $, $\alpha_{D_2} = a $, $\alpha_{D_3} = 0 $ and $\alpha_{D_4} = \frac{1}{2}$, where $a \in (0,\frac{1}{2})$. We also know the canonical functions from the previous section,
\begin{align}
\omega\left(\mathcal{AC}_k (D_1)\right) = {} & \frac{1}{X_{13} X_{14}} + \frac{1}{X_{24} X_{14}} + \frac{1}{X_{13} X_{36}} + \frac{1}{X_{24} X_{26}} +\frac{1}{X_{26} X_{36}} \ , \\
\omega\left(\mathcal{AC}_k (D_2)\right) = {}& \frac{1}{X_{13} X_{36}} + \frac{1}{X_{13} X_{15}} + \frac{1}{X_{26} X_{36}} + \frac{1}{X_{25} X_{15}} +\frac{1}{X_{26} X_{25}} \ , \\
\omega\left(\mathcal{AC}_k (D_3)\right) = {} & \frac{1}{X_{13} X_{15}} + \frac{1}{X_{13} X_{46}} + \frac{1}{X_{25} X_{15}} + \frac{1}{X_{25} X_{26}} +\frac{1}{X_{26} X_{46}} \ , \\
\omega\left(\mathcal{AC}_k (D_4)\right) ={} & \frac{1}{X_{13} X_{46}} + \frac{1}{X_{26} X_{46}} + \frac{1}{X_{26} X_{35}}  + \frac{1}{X_{13} X_{35}} \ .
\end{align}
The full amplitude for the example in section \ref{sixpoint} is then,
\begin{equation}
\widetilde{\mathcal{M}_6} = \sum_{D=1}^{4} \alpha_D \omega(\mathcal{AC}_k(D))+\text{(cyclic permutations).}
\end{equation}

\begin{equation}
    \begin{split}
     \widetilde{\mathcal{M}_6} ={} & \frac{1}{X_{13}X_{14}}+  \frac{1}{X_{13}X_{15}}+ \frac{1}{X_{13}X_{35}}+ \frac{1}{X_{13}X_{36}}+ \frac{1}{X_{13}X_{46}}+ \frac{1}{X_{14}X_{15}}+ 
     \frac{1}{X_{14}X_{24}}\\ & + \frac{1}{X_{14}X_{46}}+ \frac{1}{X_{15}X_{24}}+ \frac{1}{X_{15}X_{25}}+ \frac{1}{X_{15}X_{35}}+ \frac{1}{X_{24}X_{25}}+ \frac{1}{X_{24}X_{26}} + \frac{1}{X_{24}X_{46}}\\ & + \frac{1}{X_{25}X_{26}}  + \frac{1}{X_{25}X_{35}} + \frac{1}{X_{26}X_{35}} + \frac{1}{X_{26}X_{36}} + \frac{1}{X_{26}X_{46}} + \frac{1}{X_{35}X_{36}} + \frac{1}{X_{36}X_{46}} \ .
    \end{split}{}
\end{equation}{}

\section{Factorization of amplitudes} \label{factorization}
One of the great virtues of the positive geometries program is that unitarity and locality arise as consequences of the geometric properties of the positive geometry.  In \cite{Arkani-Hamed:2017mur} and \cite{Banerjee:2018tun}, it was shown that geometric factorization of the associahedron and Stokes polytope implied the amplitude factorization which, in turn, implied that tree-level unitarity and locality are emergent properties of the positive geometry. In this section we show that this is indeed true for general planar scalar field theories. We will first argue that the geometric factorization of accordiohedron holds and then show how this leads to the factorization of the amplitude.

Suppose we are looking at a n-point amplitude with $v_3$ three-point vertices, $v_4$ four-point vertices, and so on. Given any diagonal $(i,j)$ of the $n$-gon, consider all dissections which contain $(i,j)$ and consider all the corresponding kinematic accordiohedra.\footnote{Here the dissections are dissections of the $n$-gon into $v_3$ triangles, $v_4$ quadrilaterals, and so on.} The diagonal $(i,j)$ splits the $n$-gon into two sub-polygons. We will call these sub-polygons the left and the right sub-polygons. Let $\mathcal{AC}_k(D)$ be the kinematic accordiohedron corresponding to a given dissection $D$ containing the diagonal $(i,j)$. Now, let us approach the facet $X_{ij}$ in the kinematic accordiohedron $\mathcal{AC}_k(D)$ by taking the limit of it going on-shell. Now, this facet, the restriction $\mathcal{AC}_k(D)|_{X_{ij}}$ factorizes combinatorially into two lower point accordiohedra $\mathcal{AC}_k(D_1)$ and $\mathcal{AC}_k(D_2)$. Where $D_1$ and $D_2$ are dissections of the left and the right sub-polygons respectively. They are such that $D_{1}\cup \lbrace{X_{ij}\rbrace}\cup D_2 = D$. How this happens is as follows.

In order to show that the facet decomposes into two lower point accordiohedra, the proof given in \cite{Banerjee:2018tun} can be generalized. We simply have to establish that the planar variables associated with the diagonals of left polygon can be written in terms of those in $D_1$ and similarly for $D_2$. Hence in order to prove this assertion we need to show that any $X_{kl}$ such that $(k,l)$ is a diagonal in the left polygon can be written as linear combination of $X_{ij}$ and the diagonals of $D_1$. To do this, it may be noted as in \cite{Arkani-Hamed:2017mur} that any such planar variable $X_{ab}$ can be written as $X_{ab} = X_{ij} + \sum_{i<c<d<j}X_{cd}$. Recalling that those $X_{cd}$ for which $(c,d)$ is not in $D$ are put to constant by the conditions of the preceding section completes the proof.

Having established this, the factorization of the amplitude follows. This assertion is based on the following facts.
As accordiohedron is a positive geometry, we know that its canonical form satisfies the following properties of a positive geometry. (For details, refer \cite{Arkani-Hamed:2017mur} and \cite{Arkani-Hamed:2017tmz})

\begin{equation}
   \text{Res}_H \omega_{\mathcal{A}}= \omega_{\mathcal{B}}
\end{equation}
where $\omega_{\mathcal{A}}$ is defined on the embedding space and $H$ is any subspace in the embedding space which contains the face $\mathcal{B}$. It is also known that if $\mathcal{B}= \mathcal{B}_1 \times \mathcal{B}_2$ then 
\begin{equation}
    \omega_{\mathcal{B}}= \omega_{\mathcal{B}_1} \wedge \omega_{\mathcal{B}_2}
\end{equation}
Now it is easy to see that 
\begin{equation}
    \text{Res}_{X_{ij}=0} \Omega[\mathcal{AC}_k(D)] = \Omega[\mathcal{AC}_k(D_1)] \wedge \Omega[\mathcal{AC}_k(D_2)] \hspace{1cm} \forall D.
\end{equation}
We thus see that residue over each accordiohedron which contains a boundary $X_{ij} \rightarrow 0$ factorizes into residues over lower dimensional accordiohedra. This factorization property naturally implies factorization of the amplitude as follows. By considering all the kinematic accordiohedron associated with dissections containing the diagonal $(i,j)$, we can compute $\widetilde{\mathcal{M}_{L}}$ and $\widetilde{\mathcal{M}_{R}}$ which correspond to left and right sub-amplitudes respectively. This immediately implies that  

\begin{equation}
 \widetilde{\mathcal{M}_n}|_{X_{ij}=0} =  \widetilde{\mathcal{M}_{L}}\frac{1}{X_{ij}}\widetilde{\mathcal{M}_{R}},
\end{equation}

This proves the physical factorization of the amplitudes.

\section{Conclusion}

In this paper, we reported on the progress of incorporating polynomial scalar interactions in the ``Amplituhedron" program. Although we focused on certain lower point amplitudes in $\lambda_3\phi^{3} + \lambda_4\phi^{4}$ theories, we believe that the accordiohedron polytope is generic enough to analyze planar amplitude in generic scalar field theories with polynomial interactions. 

Our attempt to understand (tree-level and planar) amplitudes in terms of scattering forms and positive geometries, appears to have led us to a coherent picture. Certain convex realizations of accordiohedra \cite{padrol2019associahedra} in the positive regions of kinematic space are the basic building blocks for the S matrix for these theories. A weighted sum over the scattering forms produces scalar scattering amplitudes where, as we have shown, the weights are certain combinatorial data determined by the requirement that any vertex which is shared by many  accordiohedra has a sum total of unit residue . The question of determining the weights for generic accordiohedra remains an interesting and vital open problem --- although for $\phi^{p}$ interactions, impressive progress was reported in \cite{Raman:2019utu}. 

Many interesting questions remain open. A detailed analysis of ``BCFW-type" recursion relations \cite{He:2018svj} in the case of generic scalar field theories seems plausible and may shed new light on the relationship of ``residue at infinity" in the context of on-shell recursion. The relationship with CHY formulae for scalar field theories with mixed vertices is another interesting and open direction to pursue. As a generalization of Halohedron where only compatible dissections are taken into account does not appear to be available in mathematics literature --- extending the positive geometry ideas to loop integrands in generic scalar field theories remains an open problem.  We hope to come back to at least some of these issues in near future.

\subsection*{Acknowledgements}
We would like to thank Alok Laddha for significant discussions over the course of this project and for also going through the manuscript. We are thankful to Frederic Chapoton for clarifying many of our queries and pointing out reference \cite{padrol2019associahedra} to us. We would also like to thank Akavoor Manu and Debodirna Ghosh for discussions over the initial phase of this project. MJ would like to thank TIFR for the hospitality provided during the later part of this project. N.K. would like to thank Yasha Neiman for his helpful comments on this work and support during the initial stages of writing. He would also like to thank Sudip Ghosh for a helpful discussion. This work is partially supported by a grant to CMI from the Infosys Foundation.

\bibliographystyle{utphys}
\bibliography{Accordiohedron_v3}

\end{document}